\documentclass[useAMS]{mn2e}

\def\lta{\lower2pt\hbox{$\buildrel {\scriptstyle <} 
   \over {\scriptstyle\sim}$}}
\def\gta{\lower2pt\hbox{$\buildrel {\scriptstyle >} 
   \over {\scriptstyle\sim}$}}
\def\E{{\cal E}} \def\eB{\epsilon_B} \def\eBr{\epsilon_{Br}}
\def\eBf{\epsilon_{Bf}} \def\epsilon{\varepsilon}
\def\ie{i.e.}

\def\Meszaros{M\'esz\'aros }

\begin{document}

\title[Prompt Gamma-ray and Early Afterglow Emission in the External Shock Model]
{Prompt Gamma-ray and Early Afterglow Emission in the External Shock Model}
\author[E. McMahon, P. Kumar and A. Panaitescu]{E. McMahon, P. 
Kumar and A. Panaitescu  \\
        Department of Astronomy, University of Texas, Austin, TX 78712}

\maketitle
\begin{abstract}
We describe our attempt to determine if gamma-ray burst (GRB) and afterglow 
emissions could both arise in external shocks for simple GRBs --
bursts consisting of just a few peaks in their lightcurves.
We calculate peak flux and peak frequency during the gamma-ray burst for
ten well observed bursts using the same set of parameters that are
determined from modeling afterglow emissions. We find the gamma-ray
emission properties for 970508 (which had a single peak lightcurve)
fit nicely with the extrapolation of its afterglow data, and therefore
this burst was likely produced in the external shock. One can explain
two other bursts in this sample as forward shock synchrotron emission
provided that the magnetic field parameter
during the burst is close to equipartition, and larger by a factor
$\sim 10^2$ than the afterglow value at $\sim 1$day.
The remaining seven
bursts cannot be explained in the external shock model even if we allow the 
energy fraction in electrons and magnetic 
field and the density of the surrounding medium to take on 
any physically permitted 
value; the peak of the spectrum is above the cooling frequency,  
therefore the peak flux is independent of the latter of these two               parameters, and is
too small by about an order of magnitude than the observed values.
We have also considered inverse-Compton 
scattering in forward and reverse shock regions and find that it can
explain the $\gamma$-ray emission for a few bursts, but requires
the density to be 1--2 orders of magnitude larger than a typical 
Wolf-Rayet star wind and much larger than permitted by late afterglow
observations.

We have also calculated emission from the reverse shock for these ten bursts
and find the flux in the optical band for more than half of these bursts 
to be between 9th and 12th magnitude at the deceleration time 
if the reverse shock microphysics parameters are same as those found from 
afterglow modeling and the deceleration time is of order the burst duration. 
However, the cooling frequency
in the reverse shock for most of these bursts is below the optical
band, and therefore the observed flux decays rapidly with time (as $\sim
t^{-3}$) and is unobservable after a few deceleration times. It is also
possible that the deceleration time is much larger than burst duration 
in which case we expect weak reverse shock emission.
\end{abstract}

\section{Introduction}

The localization of Gamma-Ray Bursts (GRBs) and the discovery of their 
X-ray afterglows by the \emph{BeppoSAX} satellite in 1997 has greatly 
improved our understanding of GRBs over the last 7 years. These x-ray, 
optical, and radio afterglows are thought to be produced when an external 
shock heats the surrounding medium, with radiation 
being produced via synchrotron from the heated material.  We now 
also know from the  spectroscopic confirmation of the GRB 030329/SN2003dh 
connection (Matheson et al. 2003, Stanek et 
al. 2003) that at least some long duration GRBs are produced by the 
collapse of massive stars. There is, however, considerable uncertainty
surrounding the nature of the inner engine of GRBs, and we lack a 
definitive understanding for how the $\gamma$-ray emission is produced. 
This paper is an attempt to understand how $\gamma$-ray emission
is produced in GRBs.

Multiwavelength 
afterglow data 
have enabled us to do broadband modeling of late-time afterglows.  This 
broadband modeling results in the determination of burst energy, 
microphysical shock parameters, beaming angle, and environmental 
properties (surrounding density and stratification). 
Further improvements to our understanding of GRBs requires analysis/modeling
of both the GRB and afterglow together, which we undertake here.

In this paper, we use parameters we determine for 10 bursts
by modeling their broadband afterglow emissions to extrapolate  
the radiation calculation back to the burst duration, 
with the goal of 
determining whether synchrotron emission from the forward shock can 
account for both the GRB prompt emission and the late-time afterglow.
This is especially applicable for the $\sim$10\% of bursts with a single 
pulse FRED (fast rise, exponential decline) GRB light-curve, where 
a single external shock is expected to produce the emission. This 
exercise is, however, carried out for all bursts in our sample, including       those with
moderately complex GRB lightcurves.

Internal shocks were suggested as a 
mechanism for producing $\gamma$-ray emission because external
shocks are not capable of producing rapid variability seen in many GRB
lightcurves, whereas variability arises naturally in internal shock
models, reflecting fluctuations associated with the central engine
(Rees \& \Meszaros 1994; Piran, Shemi \& Narayan 1993; Katz 1994).
For GRB lightcurves consisting of a single peak or just a few peaks, 
this rationale for internal shocks does not apply and
such bursts could be produced in external shocks. 

The determination of kinetic energy release in relativistic ejecta for 
ten  bursts by modeling their broadband afterglow lightcurves 
suggests that the efficiency for $\gamma$-ray production is typically in 
excess of 50\% (Panaitescu \& Kumar, 2002 hereafter PK02). Such a high 
efficiency cannot be
achieved in internal shocks; some published claims to the contrary 
(e.g.  Beloborodov, 2000) achieved high efficiency by colliding 
shells with very large relative Lorentz factor (hereafter LF), however in 
this case the emergent spectrum peaks at energies much larger than observed 
values. External shocks, on the other hand, can very efficiently convert bulk 
kinetic energy to radiation.

In addition to the problem of efficiency for the internal shock model
we describe below other reasons for considering the external shock model for
the generation of $\gamma$-ray emission for many of the ten bursts 
we consider in this paper (table 1 lists the ten bursts).

 The 320-1090 kev light-curve for 990123 consisted of two broad peaks 
of duration $\sim 10$s each. Comparing this time scale with the deceleration
time ($t_d$) of $\lta50$s -- which is inferred from the peak of the prompt
optical emission -- suggests that $\gamma$-ray emission is produced 
within a factor 2 of the deceleration radius\footnote{When the outermost 
$\gamma$-ray producing shell undergoes deceleration and is heated by
the reverse shock it produces optical flash, and its radius increases as 
$\sim t^{1/4}$ for $t\gta t_d/4$ (where $t$ is the observer time).
Therefore, the increase in the radius for a 5-fold increase in
time is less than a factor 2.}. 
In the internal shock model for $\gamma$-ray
production, the near equality of the radius where shells collide to
produce $\gamma$-rays and the deceleration radius is a coincidence, 
 whereas in the external shock model this is what one expects. 
It should be noted that the short time scale variability seen in 
990123 (Fenimore et al. 
1999) had an amplitude of $\sim$20\% and could have arisen due to small
scale turbulence in the shocked fluid. The observed low energy spectral 
index $\alpha$ for this burst was 0.4 ($f_\nu\propto \nu^{0.4}$) whereas in
internal shock models we expect $\alpha\sim -0.5$ due to short cooling
time for electrons or low cooling frequency (Ghisellini et al. 1999).
 
The lightcurve for GRB 970508 was a FRED, 980519 was similar to a FRED,
and 000301c lightcurve was perhaps a FRED (Smith et al. 2002), however 
because of the low temporal resolution of the Ulysses observation (0.5
sec) we are 
unsure of it. One might expect these bursts to arise in an external shock.
Two other bursts in our sample of ten -- 980703 \& 991208 -- had
lightcurves consisting of two smooth peaks, and therefore are good
candidates for a possible origin in an external shock. GRBs 990510 \&
991216 lightcurves had more fluctuations than the bursts mentioned above,
however they each had two broad peaks and a number of sub-pulses
superimposed on them, and do not require internal shock to produce this
modest variability. There are no lightcurves available for the remaining
two bursts in our sample, 000418 \& 000926, which were both detected by the 
IPN. It turns out that for all of these bursts, except
970508, the simplest theoretical model of synchrotron emission in the 
forward shock fails badly to explain their $\gamma$-ray emission (\S2).
Moreover, none of the possibilities we explore in the general framework
of an external shock model seem to work satisfactorily.

In two GRBs (990123 and 021211), a bright, steeply
falling off ($\sim t^{-2}$) early optical emission was observed.  This
has been explained by radiation from the reverse shock heated
ejecta from the explosion.  We have seen this emission from only these
two bursts, while there are many cases for upper limits within a few
hundred seconds after the GRB time and even a few bursts (e.g. 030418 and
021004) with early afterglow detections that do not exhibit the
bright, steep optical decay.  In this paper, we also estimate the reverse
shock emission at deceleration for these ten bursts, and discuss
possible reasons for numerous non-detections.   

 \S2 outlines the afterglow fitting and describes our method 
for calculating the flux and the peak frequency during the GRB. A comparison 
between the theoretical calculation and $\gamma$-ray observations is also
described in \S2.
Some alternate possibilities to explain the $\gamma$-ray observations such 
as inverse Compton in the forward or the reverse shocks, pair enriched 
ejecta, or high density clumps in the circum-stellar medium,
are discussed in \S3.  In \S4, we 
discuss reverse shock optical emission at deceleration.

\section{Afterglow to $\gamma$-ray Emission}

The afterglow modeling is described in detail in Panaitescu \& Kumar (2001)
and (2002). Briefly, we determine the collimated fireball dynamics by numerical
integration of a simplified set of jet propagation equations, keeping track
of radiative loss of energy due to synchrotron and inverse-Compton emissions.
The synchrotron peak and cooling frequencies are calculated by assuming that
a certain constant fraction of the thermal energy of the shocked fluid
is imparted to electrons and magnetic field. The effect of IC loss
including the proper Klein-Nishina cross-section is included in the 
calculation of the cooling frequency. The observed lightcurves are calculated 
by integrating the emissivity over equal arrival time surface. All of the
unknown parameters, which include jet opening angle, the total 
energy release in the explosion (which is is the sum of the kinetic energy 
given in PK02 and the energy in $\gamma$-ray radiation), the fraction of 
energy in electrons ($\epsilon_{ef}$), and the fraction in magnetic field 
($\eBf$), are obtained by fitting the observed light-curves and the 
spectrum with the theoretically calculated curves by a $\chi^2$ minimization. 
The parameter $\epsilon'_{ef}$, which determines the minimum thermal Lorentz 
factor of electrons, is 0.1 for all bursts for which $p>2$
(PK02). Since the high energy spectral index during the burst gives $p>2$ we 
set $\epsilon'_{ef}=0.1$ for all bursts in our calculations during
the gamma-ray burst.

Using these parameters we estimate the frequency where the spectrum 
($\nu f_\nu$) peaks and the flux ($f_\nu$) at this peak at
deceleration (which we assume is half of 
the $\gamma$-ray burst duration). The results for ten bursts are summarized
in Table 1 for a uniform circumburst density (wind circumburst medium 
had similar results, and for brevity are not listed here). 
The theoretical results are compared with the observed data for these
bursts (see Table 1). Note that for six out of ten bursts in the
table the peak frequency during the burst is within a factor of about 2 
of the the observed value which we consider a reasonably good agreement.
However, in four of these cases the theoretical peak flux is smaller 
than the observed value by an order of magnitude or more.
For GRB 970508, which was a single peaked FRED  
burst, the fluxes are in good agreement. Therefore, for this burst the
$\gamma$-ray emission could arise in an external shock; it is highly 
encouraging to see the forward shock model works so well to explain 
observations all the way from the $\gamma$-ray emission at 10s to radio 
at 100s of days. However, for 
000301C which was also likely a single pulse FRED, 
and 980703 \& 991208 each of which 
contain two simple peaks in their $\gamma$-ray light-curve and are 
therefore good candidates for external shock mechanism for $\gamma$-ray 
production, the discrepancy between theory and observation is large.

To understand how sensitive the $\gamma$-ray emission is to errors in 
afterglow modeling and parameter determination, and to consider some 
possible solutions within the framework of the external shock model, 
we present an analytical derivation of the main results.

The forward shock synchrotron {\sl injection} frequency, $\nu_{if}$, and 
the flux at the peak of the $F_\nu$ spectrum are (Wijers \& Galama, 1999)
\begin{equation}
\nu_{if}(t)={0.98 q B\gamma_i^2\Gamma\over 2\pi m_e c(1+z)} 
\label{num}
\end{equation}
\begin{equation}
 F_{pf}(t) = {N_e P_{\nu_p}\Gamma(1+z)\over 4\pi d_L^2}
\label{fnu}
\end{equation}
where $q$ \& $m_e$ are electron charge and mass, $m_p$ is proton mass,
$d_L = 2c\sqrt{1+z}[(1+z)^{1/2} -1]/H_0$ is the luminosity distance, 
$\gamma_i=\epsilon'_{ef} (m_p/m_e) (\Gamma - 1)$ is the minimum thermal LF of
electrons (the electron distribution for $\gamma>\gamma_i$ is assumed to be
a power-law of index $p$, \ie $dN_e/d\gamma\propto \gamma^{-p}$),
$N_e=4\pi A R^{3-s} m_p^{-1}/(3-s)$ is the total number of swept-up ISM
electrons, $\rho_0=A R^{-s}$ is the density of the medium just ahead of 
the shock, $\Gamma$ is the bulk LF of shocked gas,
\begin{equation}
B = 4c\Gamma \left[ 2\pi\epsilon_{Bf} A R^{-s}\right]^{1/2}, \quad\;
     R=(4-s) c\Gamma^2 t/(1+z),
\label{B}
\end{equation}
$t$ is the observer time, and
\begin{equation}
P_{\nu_p} = {1.04 q^3 B\over m_e c^2}
\end{equation}
is the power radiated per electron per unit frequency, in the shell comoving 
frame, at the peak of the synchrotron spectrum. The numerical factors
of 1.04 in the above equation and $0.98$ in equation (\ref{num}) are
taken from Wijers and Galama (1999) for $p=2$.

The synchrotron injection frequency and peak flux, at deceleration,
for the particular cases of $s=0$ \& 2 are given below
\begin{eqnarray}
  \nu_{if}(t) = {\epsilon'_{ef}}^{2}\eBf^{1\over 2} \E_{52}^{1\over 2}
   (1+z)^{1\over 2} t_{d,1}^{-3/2}
 \hfill      
\nonumber\\
 \times \left\{   \begin{array}{ll}
    1.01\times10^{4} \;{\rm kev}  &  {\rm s=0}
 \\
    1.7\times10^{4} \;{\rm kev} &  {\rm s=2}
                                \end{array}  \right.
\label{numa}
\end{eqnarray}
\begin{equation}
F_{pf}(t) = {\eBf^{1\over 2}\E_{52}^{1\over 2} \over \left[ (1+z)^{1/2} -1\right]^2}
  \hfill \times \left\{  \begin{array}{ll}
   1.8 \E_{52}^{1\over 2} n_0^{1/2} \;{\rm mJy} & {\rm s=0}  \\
   {4.2\times10^{3} A_* \over [t_d/(1+z)]^{1/2} } \;{\rm mJy} & {\rm s=2}
                         \end{array}  \right.
\label{fnua}
\end{equation}
where $A_* = A/(5\times 10^{11})$ g cm$^{-1}$, $\E$ is the isotropic 
equivalent of energy release in the explosion, $t_d$ is the observer
frame  deceleration
time in seconds, and an integer subscript $n$ on a variable $X$, 
$X_n$, means $X/10^n$. 
In the derivation of the above equations we substituted for $\Gamma$ using
the equation $4\pi AR^{3-s} c^2 (\Gamma^2-1)/(3-s) = \E/2$ 
at deceleration which states that half of the original kinetic energy 
of the explosion ($\E/2$) is deposited into swept-up ISM; the LF at
deceleration is given by:
\begin{equation}
\Gamma_{d,2}  
   = \left\{  \begin{array}{ll}
 4.17 \left({\E_{52}\over n_0}\right)^{1/8}
              \left[{(1+z) \over t_d}\right]^{3/8} & {\rm s=0}  \\ 
 0.74 \left[{(1+z) \E_{52} \over A_* t_d}\right]^{1/4} & {\rm s=2}
                         \end{array}  \right.
\label{gam}
\end{equation}

We next calculate the electron cooling frequency. For this we need the
Compton $Y$ parameter, defined as $Y\equiv \tau_T \int d\gamma_e\,
\gamma_e^2 (dn_e/d\gamma_e)$, with $\tau_T$ being the column density of 
electrons times the Thomson cross-section. The $Y$-parameter is obtained 
by solving the equation describing the radiative loss of energy -

\begin{equation}
{d\over dt'}\left[ m_e c^2\gamma_e {dn_e\over d\gamma_e}\right] = - {dn_e\over d\gamma_e}
    {(1+Y) \sigma_T B^2 \gamma_e^2 c\over 6\pi}, 
\end{equation}
where $t'$ is the comoving time, $B$ is the magnetic field which we assume
is uniform, and $\sigma_T$ is the Thomson scattering cross-section. 
Consider the comoving frame down-stream fluid velocity to be $v$.
This relates $t'$ and the comoving radial coordinate $r'$ viz. $dr'=v\, dt'$.
Changing the independent variable from $t'$ to $r'$ and integrating the
above equation over the electron distribution we find
\begin{equation}
{d (\epsilon_e U)\over dr'} = - {(1+Y)\sigma_T B'^2 c n_e \overline{\gamma_e^2}\over
               6\pi v},
\end{equation}
where $U$ is the thermal energy density of shocked fluid, and 
$\overline{\gamma_e^2}$ is the average $\gamma_e^2$. Integration of this
equation over $r'$ for a highly relativistic shock ($v\sim c$) and
highly radiative fluid ($\nu_i \gg \nu_c$) we find

\begin{equation}
Y(1+Y)\approx {\epsilon_e\over 4\eB}.
\label{y}
\end{equation}

The calculation of the cooling LF of electrons, $\gamma_c$, at deceleration,
is straightforward and is given below; $\gamma_c$ is the LF of electrons
that lose their energy in a time available since crossing the shock front 
averaged over the population, given by $t_c\sim t_d/3^{(2-s)/2}$.

\begin{eqnarray}
\gamma_c(t_d) = {3\pi m_e c (1+z)\over 2\sigma_T B'^2 t_c \Gamma_d (1+Y)}=
  \nonumber\\
{3 (1+z)^{1-s} m_e \Gamma_d^{2s-3}\over 64\sigma_T\epsilon_B A c (4ct_d)^{s} 
   t_c (1+Y)}
\label{gamcd}
\end{eqnarray}
This in turn is used to calculate the synchrotron frequency, in observer
frame, corresponding to the LF $\gamma_c$, and is referred to as the 
cooling frequency ($\nu_c$). 

\begin{displaymath}
\nu_c(t_d) = {qB' \gamma_c^2\Gamma_d\over 2\pi m_e c (1+z)}
\end{displaymath}
\begin{equation}
 \hfill  {\rm or}\quad \nu_c(t_d) = \left\{ \begin{array}{ll}
     {3.2\over (\E_{52}t_d(1+z)\eB)^{1/2}\epsilon_e n_0} \;\quad{\rm ev} 
                          & {\rm s=0} \\
   2.2\times10^{-7}{ (\E_{52}t_d)^{1/2}\over \eB^{1/2} A_*^2\epsilon_e(1+z)^{3/2}}
        \quad{\rm ev} & {\rm s=2}
                     \end{array}  \right.
\label{nuc}
\end{equation}
where we have made use of equations (\ref{gam}), (\ref{y}), and (\ref{gamcd}).
We see that the cooling frequency at deceleration is typically much smaller
than the synchrotron injection frequency (see eq. \ref{numa}) and thus
the peak of the $\nu f_\nu$ spectrum will generally be at $\nu_{if}$. For
the case where $\nu_{if}>\nu_c$, the flux at the peak of the $\nu f_\nu$ 
spectrum is obtained by using equations (\ref{numa}), (\ref{fnua}) \& 
(\ref{nuc}) and is given here

\begin{eqnarray}
F_{\nu_p}(t_d) = {(\E_{52} t_d)^{1/2}(1+z)^{-1/2} \over 
   \epsilon'_{ef} \epsilon_{ef}^{1/2}\left[ (1+z)^{1/2} -1\right]^2}
  \nonumber\\
 \times \left\{  \begin{array}{ll}
   6\times10^{-4} \;{\rm mJy} & {\rm s=0}  \\
   3.5\times10^{-4} \;{\rm mJy} & {\rm s=2}
                         \end{array}  \right.
\label{fnub}
\end{eqnarray}
Note that the flux is independent of $\eB$ and the density of the ISM --
the two parameters that have the largest error associated with them
in afterglow modeling (PK02).
Using equations (\ref{numa}) and (\ref{fnub}) we calculate the peak
frequency and the flux during the $\gamma$-ray burst 
and the results are in good agreement with the numerical calculation
result presented in table 1. 

Table 1 shows that for all those bursts for which the observed and
theoretical peak frequencies agree within a factor of two\footnote{We
have assumed that the peak frequency for 000301c was $\sim 500$ keV,
at the higher end of the peak frequency distribution, since it was a
fairly short burst (8.4 seconds).} (6 cases altogether), the theoretically
calculated fluxes, with the exception of 970508, are too small by an order of 
magnitude compared to the observed fluxes.

We see from equation (\ref{fnub}) that the flux at the peak of $\nu f_\nu$
depends only on $\E$, which is a very well determined parameter, and
$\epsilon'_{ef}$ which is 0.1 for all the bursts with $p>2$, and therefore
there is no way to reconcile the difference between the observed flux and the 
theoretical expectation in the simplest version of the synchrotron emission 
in the external shock model. Any error in the parameter determination
from afterglow modeling does not affect our calculation of the $\gamma$-ray
flux. In other words, even if we consider $\eBf$ and the ISM density $n$
during the burst to take on values completely unrelated to what is determined
from afterglow modeling the observed flux cannot be reconciled
with the theoretical expectations in the forward shock model. 
This is a very robust result and effectively rules out synchrotron
origin for $\gamma$-ray emission in the external shock model for 
these bursts.

In four cases, viz. 980519, 980703, 990123 \& 000418, the theoretically 
calculated peak frequencies are much smaller than the observed value; 
these are the four cases with the smallest $\eBf$ as determined
by the afterglow modeling (see Table 1). Could a larger $\eBf$ 
at early times, as in the case of 021211 (Kumar \& Panaitescu, 2003, hereafter 
KP03), 
explain the peak frequencies for these four cases? If $\eBf$ were
to be $0.5$ during the burst for 980519 \& 000418 we can explain the
$\gamma$-ray emission for these GRBs. However, even if we
set $\eBf=1$ during the burst for 980703 \& 990123 the synchrotron 
frequency falls short of the observed value. Could the gamma-ray burst
in these cases arise as a result of inverse-Compton (IC) scattering of the
synchrotron radiation in the forward or the reverse shock? We consider this
possibility, and some others, in the next section. We also investigate 
whether the peak flux of the IC 
component might be able to match the observed flux for the other five bursts
which have too small synchrotron flux.

\begin{table*}
\centering
\begin{minipage}{126mm}
\caption{Forward Shock Emission at Deceleration For Homogeneous External 
Medium}
\label{fstable1}
  \begin{tabular}{cccccccl}
  \hline
 & $\E$ &  &
$\nu_p(t_d)$ & $\nu_{p,obs}$ &$f_{\nu_p}(t_d)$ & $f_{\nu_p,obs}$ & \\ Burst & ($\times10^{52}$ ergs) & $\eBf$ 
 & (keV) &(keV) & (mJy) & (mJy)& Refs.\\  
  \hline
970508 & 7.92 & $4.5\times10^{-2}$ & 98.4 & 100 &0.51  & 0.6 & 2,4 \\
980519\footnote{Redshift not known for this burst, $z=1$ used.} & 99.8 & $3.5\times10^{-5}$ & 7.0 &700 &1.25  & 0.3 & 3,4 \\
980703 & 0.40 & $3.0\times10^{-3}$ & 0.8 & 370& 0.21  & 0.3 & 3\\
990123 & 277.5 & $9.9\times10^{-4}$ & 6.6 &780 &1.24  & 9.0 & 1,2,3,5\\
990510 & 55.5 & $5.0\times10^{-3}$ & 184.9 &160 &0.27  & 4.0 & 1,2,3\\
991208 & 11.9 & $3.8\times10^{-2}$ & 118.7 &190 &0.78  & 20.0 & 1,3,6\\
991216 & 100.0 & $6.2\times10^{-3}$ & 153.4 &410 &1.0  & 30.0 & 1,3\\
000301c & 8.19 & $1.0\times10^{-1}$ & 541.6 &$500$\footnote{No $\nu_{p,obs}$ available for this burst; a value of 500 kev assumed.}&0.04 & 0.3 & 1,3,7\\
000418 & 8.05 & $1.6\times10^{-2}$ & 25.2 &280 &0.32  & 0.4 & 1,3\\
000926 & 39.7 & $8.1\times10^{-2}$& 198.4 &130 &0.21  &0.7 & 1\\
\hline
\end{tabular}
\medskip

$\epsilon'_{ef} = 0.1$, $\epsilon_{ef} = 0.5$ for all bursts, 
$n_0$, $\theta_0$ are equal to values found from afterglow modeling.
\medskip

References: (1) Mazets (personal communication); (2) Amati, L. et al. 2002; (3) Jiminez, R. et al. 2001; (4) Nicastro, L. et al 1999; (5) Briggs, M. S. et al 1999; (6) Hurley, K. et al 2000; (7) Smith, D. A. et al 2002
\end{minipage}
\end{table*}

\section{Gamma-rays in external shock: some alternate possibilities}

We consider below (\S3.1) a combination of
synchrotron and inverse-Compton processes in the forward and reverse shocks
to determine if this could explain the $\gamma$-ray emission properties
for the nine ``problem bursts'' in our sample of ten discussed in the
last section. In \S3.2 we discuss if a collision between the GRB ejecta
and a high density clump might be able to explain the large $\gamma$-ray flux
for the five of the bursts in our sample, and in \S3.3 we look into the
effect of electron-positron pair loaded ejecta on $\gamma$-ray emission. 

\subsection{Inverse Compton in the external shock}

We investigate the effect of IC in external shocks -- forward as well as
the reverse shock -- to see if the observed $\gamma$-ray emission for
the bursts in our sample could be explained by the IC process.  

Consider the flux at the peak of the synchrotron radiation $\nu
f_{\nu}$ spectrum, $\nu_p$, to 
be $f_{\nu_p}$. We consider inverse-Compton scattering by a population of
electrons that could be distinct from the population that gives rise
to the synchrotron radiation. For instance, the synchrotron emission
could be produced in the reverse shock (RS) and the IC scattering in
the FS. Let us take the minimum thermal LF of electrons in the IC-scattering
region to be $\gamma_{min}=\min(\gamma_i, \gamma_c)$, the electron distribution
to have a break at $\gamma_b=\max(\gamma_i, \gamma_c)$ such that for 
$\gamma>\gamma_b$ the electron distribution is proportional to $\gamma^{-p-1}$;
$\gamma_i$ \& $\gamma_c$ are the injection and cooling LFs for electrons.
The peak of the IC radiation (for $\nu f_\nu$) is at

\begin{equation}
\nu_p^{IC} \sim \nu_p \gamma_b^2.
\end{equation}
If the optical depth of the medium to Thomson scattering
is $\tau_T$, then the flux at the peak for the case where $\gamma_i \ll
\gamma_c$ is 

\begin{equation}
f_{\nu_p}^{IC} \sim \tau_T (\gamma_{min}/\gamma_b)^{p-1} f_{\nu_p}.
\end{equation}
For $\gamma_c < \gamma_i$, there is a slightly different relationship.

The optical depth at deceleration in the forward shock is given by

\begin{equation}
\tau_T = {\sigma_T\E\over 4\pi m_p c^2 \Gamma_0^2 R_d^2},
\end{equation}
where $R_d=(4-s) c t_d \Gamma_d^2/(1+z)$ is the deceleration radius,
$\Gamma_d$ is the LF at deceleration, and $\Gamma_0\sim 1.5\Gamma_d$ is
the initial LF of the ejecta (see eq. \ref{epr}). Using
equation (\ref{gam}) this can be rewritten as follows

\begin{equation}
\tau_T  
  \hfill = \left\{  \begin{array}{ll}
 3.9\times10^{-9} n_0^{3/4} \E_{52}^{1/4} [t_d/(1+z)]^{1/4} & {\rm s=0}  \\ 
 8.7\times10^{-4} A_*^{3/2} \E_{52}^{-1/2} [(1+z)/t_d]^{1/2} & {\rm s=2}
                         \end{array}  \right.
\label{tauf}
\end{equation}
We see that the optical depth for a uniform density ISM is very small,
and therefore the inverse Compton flux due to scattering in the forward shock
region, for $s=0$, is likely to be too small to be observationally 
interesting.

The optical depth to Thomson scattering of the ejecta at deceleration
can be obtained directly from equation (\ref{tauf}) by recognizing that
the mass of the ejecta is larger than the swept-up ISM mass by a factor 
$\Gamma_d$. Thus,

\begin{equation}
\tau_T      
  \hfill = \left\{  \begin{array}{ll}
 1.5\times10^{-6} n_0^{5/8} \E_{52}^{3/8} [t_d/(1+z)]^{-1/8} & {\rm s=0}  \\
 5.4\times10^{-2} A_*^{5/4} \E_{52}^{-1/4} [(1+z)/t_d]^{3/4} & {\rm s=2}      
                         \end{array}  \right.
\label{taur}             
\end{equation}

\subsubsection{Reverse Shock Break Frequencies and Peak Flux}

To complete the calculation of inverse Compton scattering of synchrotron 
emission produced in the reverse shock (RS) region we provide below 
the synchrotron characteristic frequency and the flux from the RS
(see KP03 for details). 

The thermal energy per proton in the RS at deceleration, $e_p$, can be 
calculated using the following pair of equations

\begin{equation}
e_p = {1\over 2} \left( {\Gamma_d\over \Gamma_0} + {\Gamma_0\over \Gamma_d}\right), 
    \quad
{\Gamma_d\over \Gamma_0} = \left[ 1 + 2\left( {n_0\Gamma_0^2\over n_{ej}}\right)^{1/2}
        \right]^{-1/2}.
\label{epr}
\end{equation}
which fits the results of numerical calculations to better than 8\%
in the Newtonian, Relativistic and intermediate regimes; where $n_{ej}$
is the comoving density of the ejecta, $n_0$ is the ISM density, and 
$\Gamma_0$ is the initial LF of the ejecta.

It can be shown that $n_{ej}/n_0$ at the time when the reverse shock 
arrives at the back end
of the ejecta (which is approximately equal to the deceleration
time for the ejecta) is $1.5 \Gamma_0^2$ for a uniform density ISM and
$3.5 \Gamma_0^2$ for $s=2$ medium. The reverse shock in this case is
neither highly relativistic nor Newtonian. Using the above equation 
we find the thermal energy per proton in the RS in this case to be
0.13$m_p c^2$ ($0.067 m_p c^2$) for a $s=0$ ($s=2$) medium.

The injection frequency at deceleration for RS is smaller than the FS 
by a factor of  $\Gamma_d^2/0.13^2$ for uniform ISM and is given below 

\begin{equation}
\nu_{ir}\left( t_d \right) = \frac{0.07 q m_p^{5/2} \epsilon_{Br}^{1/2} 
  \epsilon_{er}'^{2} n_0^{1/2} R_d^{-s/2} \Gamma_d^2}
  {\left(2 \pi \right)^{1/2} m_e^3 (1+z)}.
\end{equation}

\noindent This equation, and a similar one for $s=2$, can be rewritten as     
\begin{displaymath}
\nu_{ir}\left( t_d \right) = \epsilon_{Br}^{1/2} \epsilon_{er}'^2  
\end{displaymath}
\begin{equation}
 \times    
    \left\{   \begin{array}{ll}
  37\, n_0^{1/4} \E_{52}^{1/4} (1+z)^{-1/4} t_d^{-3/4} \;{\rm ev} & {\rm s=0}
 \\
    610\, A_*^{1/2} t_d^{-1} \;{\rm ev} &  {\rm s=2}
                                \end{array}  \right.
\label{nuir}
\end{equation}

\noindent where we have used equation (\ref{gam}) to eliminate
$\Gamma_d$.   The 
cooling frequency in the reverse shock region, when $\nu_{ir}>\nu_{cr}$, 
is given by equation (\ref{nuc})
with appropriate values of $\epsilon_B$ and $\epsilon'_e$ for the reverse
shock.  However, the reverse shock $\nu_{cr}$ is typically larger than
$\nu_{ir}$ (see Table 2), so equation (\ref{y}) is not a valid
approximation for the Compton Y parameter any longer and we must also 
use an appropriately modified form version of equation (\ref{nuc}).

The flux at the peak of the reverse shock $f_{\nu}$ spectrum
at deceleration
is larger than the peak flux from the FS by a factor $\Gamma_d$ and can be 
written as
\begin{equation}
F_{pr}\left( t_d \right) = \frac{ \left( 3 \epsilon_{Br} A \right)^{1/2} q^3 \E (1+z)^{s/2}}{m_e m_p c^3 d_L'^2 \Gamma_d^{s-1} \left(4 c t_d\right)^{s/2} \left[\left(3-s\right) \pi \right]^{1/2}} 
\end{equation}

\noindent or 

\begin{displaymath}
F_{pr}\left( t_d \right) =  {\epsilon_{Br}^{1/2} \over
   [(1+z)^{1/2}-1]^2} 
\end{displaymath}
\begin{equation}
\times  \left\{   \begin{array}{ll}
   6.8\times10^{2} (1+z)^{3/8} \E_{52}^{9/8} n_0^{3/8} t_d^{-3/8} \;{\rm mJy}  &  {\rm s=0}
 \\
   2.7\times10^{5} (1+z)^{3/4} \E_{52}^{3/4} A_*^{3/4} t_d^{-3/4} \;{\rm mJy} &  {\rm s=2}
                                \end{array}  \right.
\label{fnur}
\end{equation}
The self-absorption frequency in the reverse shock region,
$\nu_{Ar}$, is often
as large as the cooling and the injection frequencies, and therefore should be
taken into consideration in the calculation of observed flux. The 
self-absorption and cooling frequencies and the Compton Y parameter
need to be calculated together in 
a self-consistent way (as we do for all of our numerical calculations). 
However, when electron cooling is dominated by 
the inverse-Compton scattering  
and $\max\{\nu_{ir}, \nu_{cr}\}>\nu_{Ar}$, the calculation of 
self-absorption frequency is considerably simplified and is given by (KP03)

\begin{displaymath}
\nu_{Ar}\left( t_d \right) \left(\frac{\nu_{ir}}{\nu_{Ar}}
  \right)^{\alpha/2} =\left(\frac{ \epsilon_{Br}}{ {\epsilon_{er}'}
  }\right)^{1/2} 
\end{displaymath}
\begin{equation} 
   \times  \left\{   \begin{array}{ll}
   8.4\times10^{-2} \E_{52}^{3/8} n_0^{3/8} t_d^{-5/8}(1+z)^{-3/8} 
       \;{\rm ev}  &  {\rm s=0} \\
   49\, A_*^{3/4} t_d^{-1} \;{\rm ev} &  {\rm s=2}
                                \end{array}  \right.
\end{equation}

\noindent where $\alpha$ depends on the ordering of $\nu_{ir}$,
 $\nu_c$ and $\nu_{Ar}$, and is equal to $1/3$ if 
$\nu_c>\nu_{ir}>\nu_{Ar}$, and $-p/2$ when $\nu_c>\nu_{Ar}>\nu_{ir}$.

\subsubsection{Inverse Compton results}

The peak frequency and flux for the IC radiation
is calculated as described in \S3.1. 
 The calculation
of the synchrotron injection frequency is straightforward \& is carried
out as described in \S2 for the FS and \S3.1.1 for the RS. The synchrotron
self-absorption frequency is typically small in the FS and is unimportant
for IC calculation. However, in the RS the self-absorption can be larger
than the cooling frequency and these frequencies must be calculated
self-consistently; we calculate these frequencies numerically.

There are four cases of the IC scattering to consider: synchrotron in
the FS and IC in either the FS or the RS, synchrotron in RS and IC in
the RS or the FS.
We have investigated these cases numerically, and we have explored 
the parameter space -- $\E$, $\epsilon_{Br}$, $\epsilon'_{er}$, $
\epsilon_{Bf}$, $\epsilon'_{ef}$, $n$ -- for each burst to determine 
if the observed $\gamma$-ray peak frequency and flux could be explained by the 
IC radiation, either for
a uniform density circum-burst medium or a wind-like medium ($\rho\propto
R^{-2}$).
The results for each burst are described below.

For a uniform density ISM the synchrotron-IC mechanism in external shock
offers a vanishing parameter space that is consistent with the gamma-ray 
emission properties for GRB 990123. However, for a $s=2$ medium we find some
solutions where the synchrotron emission produced in the forward shock
undergoes inverse-Compton scattering in the reverse-shock region. The
density required for these solutions is $\sim 10^2$ times that normally
associated with Wolf-Rayet star winds, and much greater than what is found 
from modeling of early and late time afterglow observations. Other 
parameters such
as the energy in the explosion, micro-physics shock parameters in the
forward and the reverse shock are roughly consistent with the afterglow
observations. However, the low energy spectral index for the $\gamma$-ray
spectrum ($\alpha$) is -0.5 whereas the observed index is 0.4. Therefore,
we do not have a fully self consistent solution for the gamma-ray 
emission properties for 990123 in the external shock model. This is 
surprising in the light of the arguments for external shock (see \S1)
for this burst, and perhaps suggests that there may be some other mechanism
producing the $\gamma$-ray photons that is completely different from
the standard internal/external shocks model. In the next sub-section
we explore if 
high density gas near the deceleration radius could
explain the $\gamma$-ray emission.

In the case of 980519 there are IC solutions for $s=2$ medium where the
synchrotron radiation is produced in the forward shock \& IC in the
reverse-shock, and the parameter space consists of $A_*$ in the range of 
15 to 100 which is at least a factor of a few larger than the value of 
$A_*\sim 3.5$ determined from afterglow modeling. However, the IC solution
requires $\epsilon_B\sim 10^{-4}$ in the forward shock that is much
smaller than the value of 0.1 we find for this burst from the afterglow 
fitting when $s=2$. The RS optical peak flux of $\sim 11$th magnitude is perhaps 
not a problem for this solution.

For 980703 there are solutions found for $s=0$. However, these solutions
require $\E>10^{55}$ erg and the RS optical flux is larger than 
1 Jy, and therefore these are not acceptable solutions. The solutions
we find for this burst with $s=2$, which involve synchrotron in the FS
and IC in the RS, require $A_*\gta 5$ and other parameters are roughly 
consistent with
the values we find from the afterglow modeling; the optical flux from 
the RS is $\sim 20$ mJy. If we ignore the somewhat high density requirement, 
this burst could perhaps be produced as IC in the external shock.

For 990510 no solution is found
that is in agreement with the observed properties of this burst. The
same is true for 991208, 991216. To be precise, there are
no solutions found when the density of the medium is taken to be uniform.
 However, for a pre-ejected
wind medium there are regions in the multidimensional parameter space 
$({\E}, A_*, \epsilon_B, \epsilon_e')$ that give gamma-ray flux
and peak frequency in agreement with observations for these bursts
where synchrotron emission is produced in the forward shock and
the inverse-Compton scattering takes place in the reverse shock region.
 The problem is that for all of these ``solutions'' $A_*$
is greater than about $10^2$ which is larger than what we obtain from 
afterglow modeling by two orders of magnitude, and too large for 
winds from Wolf-Rayet stars.
Moreover, the large $\eBr$ in the reverse shock for these solutions
gives rise to optical R-band flux of about 10 Jy, or 6th magnitude, 
which is unlikely to have gone unnoticed.
Therefore we do not consider these solutions physically acceptable.

We find two different IC ``solutions" for 000418 for a $s=2$ circum-burst 
medium; one of which is synchrotron in the FS and IC in the RS, and the other 
is synchrotron in the RS and IC in the FS. 
The first scenario requires $A_*\gta 25$ (which is larger by factor 
$\sim 10^2$ than determined from afterglow observations) and $\eBf$ is
between $\sim10^{-5}$ and $2\times10^{-4}$ which is smaller by two orders of 
magnitude than the value we find from afterglow modeling. The peak optical 
flux from the RS is between 20 mJy and 3 Jy; the lower flux value in 
this case certainly poses no difficulty with observations. The latter
scenario requires $A_*\gta 400$ and $\eBr\sim 10^{-5}$ which are unlikely 
to be realized in nature.

There are also two solutions for 000301c with $s=2$.  For FS synchrotron \&
IC scattering in the RS, $A_*$ is about an order of magnitude larger than 
that found by afterglow modeling, but the optical flux from RS is not a problem, 
being $\lta 0.5$ Jy. The magnetic field parameter in the forward shock
required for the IC solution is, however, several orders of magnitude 
smaller compared with the value we find from afterglow modeling in a wind-like
medium for this burst. This together with the required high density for the CBM
makes this solution unacceptable. For the case of RS synchrotron 
\& IC scattering in the FS, the allowed parameter space to explain the
$\gamma$-ray observations require the energy to be a factor of 10 smaller
than the observed value, and $A_* \gta 100$.
The magnetic field parameter is also 100 times smaller than the
afterglow value, so we do not consider these solutions viable. 
For a uniform density medium the IC solution for 000301c requires
$\E>10^{55}$ which is more than two orders of magnitude larger than the 
energy determined from either the $\gamma$-ray fluence or the afterglow
emission. The situation for 000926 is a bit worse than 000301c.

We also note that the IC flux for 970508, in any of the four possible
combinations, at the observed peak is too small to be a significant
contributor to the observed flux. 

\subsection{Effect of density clumps in the ISM on gamma-ray flux}

In this subsection we investigate whether a dense clump of
gas in the circum-burst medium (CBM)  might
increase the flux in the $\gamma$-ray band and thereby explain flux
observations at
the peak of the $\nu f_{\nu}$ spectrum for some of the bursts in our sample.

Consider a dense clump of angular size greater than $\Gamma_0^{-1}$, and
proton number density $n$; $\Gamma_0$ is the initial Lorentz factor (LF) of
the ejecta which we assume does not decrease until it runs into the clump.
For the calculation of radiation
such a clump can be treated as a spherical object.
We take the external density to be sufficiently high that the
forward shock LF is
less than $\Gamma_0$ and the reverse shock is relativistic;
the parameter $\xi\equiv \Gamma_0^2 n/n_{ej} > 1$ determines the thermal LF
of protons in the forward and the reverse shocks; $n_{ej}$ is the
density of the ejecta when it hits the clump.
For large $n$ the thermal LF of protons in the FS is $\gamma_{p,f}=\Gamma_0
\xi^{-1/4}/2^{1/2}$ and in the RS it is $\xi^{1/4}/2^{1/2}$.

The thermal energy density in these shock regions is $4 n m_p c^2\gamma_{p,f}^2
\sim n\Gamma_0^2 m_p c^2 \xi^{-1/2}$, and therefore the magnetic field
strength $B \propto (\eBf n)^{1/2} \Gamma_0 \xi^{-1/4}$.
The synchrotron injection frequency in the forward shock is $\nu_i \propto
{\epsilon'_{ef}}^2 (\eBf n)^{1/2} \Gamma_0^4 \xi^{-1}$.

The number of swept up protons at deceleration is obtained by equating the
thermal energy of protons (in lab frame) with half the energy in the explosion
i.e.,
$N_p m_p c^2 \gamma_{p,f}^2 = \E/2$ or $N_p = \E \xi^{1/2}/(m_p c^2 \Gamma_0^2)$.

Let us assume that the distance of the clump from the center of the explosion
is $R_c$ and the forward shock travels a distance of $\delta R_c$
before the shocked material
acquires half the energy of the explosion. Therefore, $N_p = 4\pi n R_c^3
\delta = \E \xi^{1/2}/(m_p c^2 \Gamma_0^2)$. The GRB duration (in the
observer frame), if the
$\gamma$-rays are produced due the ejecta colliding with the dense
clump, is $t_\gamma \sim R_c/(c\gamma_{p,f}^2)\sim
R_c\xi^{1/2}/(c\Gamma_0^2)$. Combining this with the equation for
$N_p$ we find:  $n\Gamma_0^8\sim \xi^2 \E/(4\pi m_p c^5 t_\gamma^3\delta)$.
Substituting this back into the equation for injection frequency we find
$\nu_i\propto \eBf^{1/2}{\epsilon'_{ef}}^2 \E^{1/2} t_\gamma^{-3/2}
\delta^{-1/2}$.

The time scale in the lab frame for electrons to cool
in the forward shock is
$\sim \delta R_c/(c\gamma_{p,f})$, from which we calculate the cooling
frequency to be $\nu_c \propto (n\eBf)^{-3/2} (1+Y)^{-2}
(\delta R_c)^{-2} \propto \eBf^{-3/2} (t_\gamma \E)^{-1/2} n^{-1}
\delta^{-3/2}(1+Y)^{-2}$.

The flux at the peak of the synchrotron spectrum is $f_p\propto
\eBf^{1/2} n^{1/2} \E$. For $\nu_c<\nu_{if}$, expected for
a high density clump, the Compton $Y$-parameter is $\sim
(\epsilon_{ef}/\eBf)^{1/2}$, and the flux at the peak of $\nu f_\nu$
i.e. at $\nu_{if}$ is proportional to $\E^{1/2} t_{\gamma}^{1/2}\delta^{-1/2}
\epsilon_{ef}^{-1/2} {\epsilon'_{ef}}^{-1}$. Note a weak dependence of
the peak flux on $\delta\equiv\delta R_c/R_c$.

The distance $\delta R_c$ the FS moves before the GRB ejecta is decelerated
is obtained by calculating the density of the ejecta at $R_c$;
$n_{ej}\sim \E/[4\pi R_c^2 m_p c^2 \Gamma_0^2\max(t_* c, R_c/\Gamma_0^2)]$,
where $t_*$ is the duration of the central engine in the lab frame.
Using this expression and the definition of $\xi$ we find that 
$\delta \sim \max(t_*/t_\gamma, \xi^{-1/2})$.

Since the peak flux is proportional to $\delta^{-1/2}$, a factor of 
ten increase in the flux requires $\xi\gta 10^4$ or the clump density 
$n\sim n_{ej} (10^{-4}\Gamma_0^2)$. Therefore, for $\Gamma_0\sim 10^2$,
$n$ is of order $n_{ej}$ and we find that in order to explain the gamma-ray
flux in the clump-ejecta collision the density of the clump needs to be
similar to the ejecta as in the internal shock model! Very bright
early afterglow will be produced in such a collision which might
pose a problem for this scenario. We note that
the increase in the flux is accompanied by an increase in the peak
frequency, both of which are proportional to $\delta^{-1/2}$, but the
latter quantity can be easily adjusted by a decrease in $\eBf$
or $\epsilon_e'$ to match the observations.

\subsection{Effect of $e^{\pm}$ pairs present in the ejecta}

In this subsection we consider whether $e^{\pm}$ pairs present in the
ejecta can make  
inverse-Compton radiation match the peak flux and frequency at
deceleration.
Pairs would soften the reverse shock
spectrum, and the larger optical depth could increase IC flux at the peak.  
Adding $N_{\pm}$ pairs per proton to the ejecta increases the
number of radiating particles in the reverse shock region thus
lowering the energy per particle. This decreases $\nu_{ir}$ by
a factor of $N_{\pm}^2$ while increasing the peak flux of the
$f_{\nu}$ spectrum by a factor of $N_{\pm}$.  The cooling frequency for
the highly radiative regime is not affected, nor is it changed when the
Compton-Y parameter is much less than one; it is, however, affected
when $\nu_{ir} < \nu_{cr}$ (more common in the reverse shock
emission than $\nu_{ir} > \nu_{cr}$, see Table 2) and $Y>1$, 
increasing it by a factor $N_{\pm}^{2(p-2)/(4-p)}$.  

The flux at the cooling frequency, which is the peak of the reverse 
shock synchrotron $\nu f_{\nu}$ spectrum, is proportional to
 $N_{\pm}^{3(2-p)/(4-p)}$.
We see that for $p=2$ the cooling frequency and the peak flux are independent
of $N_\pm$. The peak flux increases by a factor of $N_{\pm}^{3/5}$ for $p=1.5$
(decreasing for $p>2$).  For inverse-Compton scattering of these reverse shock
synchrotron photons by the ejecta, the IC peak, $\nu_{p,IC} =
\gamma_c^2 \nu_c$, will increase by a factor of
$N_{\pm}^{4(p-2)/(4-p)}$.  We see again that $\nu_{p,IC}$ is not
affected when $p=2$, but can go up ($p>2$) or down ($p<2$) for other
values.  The IC flux at the peak changes
by a factor of $N_{\pm}^{6(2-p)/(4-p)}$ (from its value without
pairs present), which, similarly, does
not change much for values of $p$ around 2.  The IC peak flux
decreases for $p > 2$ and at most, can be
increased by a factor of $\sim N_{\pm}$ when the electron distribution
is very hard, i.e. $p \lta 1.5$. 

It can be shown that the peak flux for synchrotron produced in the
FS and IC in the slow cooling RS (with $Y \gg 1$) is proportional to
$N_{\pm}^{3(2-p)/(4-p)}$, decreasing for $p > 2$. If the RS was in the highly
radiative regime, the IC flux would increase as $N_{\pm}^2$; however,
the injection frequency decreases rapidly as $N_{\pm}^{-2}$, whereas
$\nu_{cr}$ is independent of $N_{\pm}$ in the fast cooling regime. 
We therefore
expect $\nu_{ir}$ to become less than $\nu_{cr}$ as $N_{\pm}$ becomes
larger than a certain value and the slow cooling regime considered
earlier once again applies.  The reverse
process -- synchrotron in the RS and IC in the FS -- has also lower
peak flux for $p>2$. Thus, we find that pairs
present in the ejecta are not likely to be able to account for the
theoretical IC and observed $\gamma$-ray flux difference.

\section{Early Afterglow Emission}

It is generally believed that the steeply falling off early afterglow 
emission observed 
from GRBs 990123 and 021211 was produced by the reverse shock heated 
ejecta from the explosion.  This emission falls off roughly as 
$t^{-1.7}$ and flux falls below the forward shock emission level after 
about $10 - 20$ minutes.  
We use the equations in \S3.3 to calculate the observed
flux in the optical R-band at deceleration for this sample of 10
bursts.  For the case 
of $\nu_{ir}<\nu_{cr}<\nu_R$, 
the flux at $\nu_R\sim 2$ eV, the R-band in observer frame, is given by

\begin{displaymath}
f_{R}\left( t_d \right) =  {\epsilon_{Br}^{{p\over4}}{\epsilon'_{er}}^{p-1}
    \epsilon_{er}^{-{1\over2}} \over [(1+z)^{1/2}-1]^2} 
\end{displaymath}
\begin{equation}
\times  
    \left\{   \begin{array}{ll}
   400\times3.3^{{p-1\over 2}} (1+z)^{{2-p\over 8}} \E_{52}^{{p+6\over 8}} 
    n_0^{{p-2\over 8}} t_{d,1}^{-{3p+2\over 8}} \;{\rm mJy}  &  {\rm s=0}
 \\
   584\times30^{{p-1\over 2}} \E_{52} A_*^{{p-2\over 4}} t_{d,1}^{-{p\over 2}} 
      \;{\rm mJy} &  {\rm s=2}
                                \end{array}  \right.
\label{fnuopta}
\end{equation}
Whereas for the case where $\nu_{ir}<\nu_R<\nu_{cr}$ the flux is

\begin{displaymath}
f_{R}\left( t_d \right) =  {\eBr^{{p+1\over4}}{\epsilon'_{er}}^{p-1}
    \over [(1+z)^{1/2}-1]^2} 
\end{displaymath}
\begin{equation}
\times  
    \left\{   \begin{array}{ll}
   290\times3.3^{{p-1\over 2}} (1+z)^{{4-p\over 8}} \E_{52}^{{p+8\over 8}} 
    n_0^{{p+2\over 8}} t_{d,1}^{-{3p\over 8}} \;{\rm mJy}  &  {\rm s=0}
 \\
   4.8{\rm x}10^4\times30^{{p-1\over 2}} \E_{52}^{3\over 4} A_*^{{p+2\over 4}} 
    t_{d,1}^{-{2p+1\over 4}} (1+z)^{-{3\over4}} 
      \;{\rm mJy} &  {\rm s=2}
                                \end{array}  \right.
\label{fnuoptb}
\end{equation}

In Table 2, we provide the theoretical estimations of the magnitude of the
 reverse shock emission for the ten bursts in our sample (for a homogeneous
external medium and assuming the RS parameters to be same as the FS),
to determine if these bursts would have had a bright optical flash. 
These results
were obtained numerically using an accurate calculation of the cooling
and the self-absorption frequencies, which can also be found in Table 2,
and the flux is found to be consistent with the analytical estimate 
given above. Our fluxes are somewhat smaller than reported in 
Soderberg \& Ramirez-Ruiz (2002). The difference is perhaps because the 
RS falls in a regime that is neither Newtonian nor relativistic where 
the usual asymptotic approximations are not very accurate, and one needs a
more accurate calculation for this intermediate case (Nakar \& Piran 2004 
have made a similar point).

\begin{table}
\centering 
\begin{minipage}{80mm}
\caption{Predicted reverse shock flux using afterglow parameters 
for homogeneous external medium}
\label{rstable1}
\begin{tabular}{ccccc}
\hline
 & & $\nu_{ir}$   &
$\nu_{cr}$ &
$\nu_{ar}$  \\ 
Burst&$R$\footnote{Reverse shock R-band magnitude 
at deceleration with parameters determined from afterglow modeling.} & 
(eV) &(eV) &(eV)\\ 
\hline 
970508 & 9.8&1.0$\times 10^{-2}$& 5.8  &$3.9\times10^{-2}$ \\
980519\footnote{Redshift not known for this burst, $z=1$ used.} 
& 17.1 &$3.8\times10^{-4}$&$5.1\times10^{5}$&$2.8\times10^{-3}$ \\
980703 & 17.4 &$2.4\times10^{-5}$&$6.0\times10^{1}$&$2.0\times10^{-3}$ \\
990123 & 14.4&5.3$\times10^{-4}$&$5.1\times10^{4}$&$2.1\times10^{-3}$ \\
990510 & 11.9&3.1$\times10^{-4}$& $3.2\times10^{1}$ & 1.2$\times10^{-2}$ \\
991208 & 8.8&1.0$\times10^{-2}$&2.0$\times10^{-2}$&$1.0\times10^{-1}$ \\
991216 & 8.5&$1.7\times10^{-4}$&3.0$\times10^{-2}$&$7.0\times10^{-2}$\\
000301c& 10.8&2.0$\times10^{-2}$&7.0$\times10^{-3}$&1.2$\times10^{-1}$ \\
000418 & 10.9&5.0$\times10^{-3}$&7.9$\times10^{-1}$&$4.5\times10^{-2}$ \\
000926 & 9.5&3.5$\times10^{-2}$&1.5$\times10^{-1}$&$1.4\times10^{-1}$\\
\hline
\end{tabular}
\end{minipage}
\end{table}

We see that for six of the ten bursts the optical flux is between 8.5 and
11-th magnitude, and the cooling frequency at deceleration is small (less
than or of order the R-band frequency).
With the cooling frequency either below or dropping below the optical 
rather quickly, and no electrons left in the RS capable of radiating
in the optical band, we would not see the expected $\sim 1/t^2$ falloff, 
but a more rapid falloff of $\sim 1/t^3$ (Kumar \& Panaitescu, 
2000). This falloff is fast, but even so, some of these bright optical 
transients could be seen for a few hundred seconds by rapid followup 
observations with a limiting magnitude of $R\sim 15$.

The remaining four have much fainter
optical emission and large cooling frequency.  The magnetic field
parameter determined from afterglow modeling for these bursts is much
lower than those with bright reverse shock emission.  With the high
cooling frequency (at least a factor of 10 above the observing band
frequency of 2 eV), we expect this emission to exhibit $t^{-2}$ falloff;
however, this faint emission may be hidden by brighter forward shock
afterglow emission.  

For those bursts which are fit equally well in an $s=2$ medium (970508, 
000418, 991208, 000301c, 991216), the reverse shock flux at deceleration 
has been calculated using the afterglow parameters in PK02. We find that 
the flux is typically a few times larger in the s=2 model than uniform 
ISM case discussed above.

According to table 2, the peak $R$ magnitude for the reverse shock 
emission for 990123 is 14.4 (using the forward shock values for all parameters 
in the reverse shock) whereas the observed peak flux was $R = 8.9$ 
(Akerlof et al. 1999). This suggests, as has been pointed out by
Zhang et al. 2003, that $\epsilon_{B}$ was larger in the RS by a factor of
about $10^2$ than in the FS.
Since $\nu_c \propto \epsilon_B^{-3/2}$, by making 
$\epsilon_{Br} = 0.07$, the cooling frequency has been lowered from 51 keV 
to 100 eV.  The R band is at about 2 eV, so the cooling frequency is still 
well above the optical at deceleration, allowing for the $\sim 1/t^2$ 
falloff that was observed. 

Bursts with small optical flux at deceleration have small $\epsilon_B$
and/or $\epsilon_e'$ in the RS and their cooling 
frequency is generally high, allowing the optical emission falling off 
as  $\sim 1/t^2$ to occur. If $\epsilon_B$ in the RS for these bursts
were larger than the FS, as found for 990123 (Zhang et al. 2003) \&
021211 (KP03), the flux will be boosted to levels comparable to that 
of the other brighter bursts. 

Besides 990123 and 021211 there have been no observations of a
bright and quickly fading early afterglow for any other bursts. 
There have been some early (time since onset of burst $< 0.01$ day or 
$\sim 10$ minutes) detections in the optical, for example 030329 and 021004, 
but the emission was not falling off as $\sim 1/t^2$.  There have been about 
18 bursts with upper limits published in the GCN Circulars (Barthelmy et 
al. 1995) and in the literature (e.g. Akerlof et al. 2000, Kehoe et al. 2001), 
for emission 
between the GRB time and 0.01 day (9 of these bursts reported in the GCNs 
had later optical afterglow detections).  Searching the GCN Circulars with 
the GRBlog website (Quimby et al. 2003), we find that the burst upper limits 
range from $R \sim 10$ for 030115 (Castro-Tirado et al. GCN 1826) at early 
times to $R \sim 20$ for XRF 030723 (Smith et al. GCN 2338) closer to 
0.01 day.  None of the bursts from the sample in this paper have upper 
limits available.  However, if the bursts with available upper limits 
are representative of the total GRB population, then it is possible that 
the bursts in our sample would have had similar upper limits, i.e., 
roughly 14-15th magnitude at $\sim 500$s. So there is a
disagreement between the theoretical expectation and the observational
upper limit.

There are several possible resolutions for this apparent discrepancy.
The small optical flux could be due to much smaller magnetic field in 
the RS compared with the FS; for 990123 and 021211, however, $\epsilon_B$
in the RS was inferred to be larger than the FS, which perhaps might not be 
the common situation.
Another possibility is that the deceleration time of GRB fireball is
of order an hour instead of the burst duration of a few tens of seconds
(assumed for the GRBs considered in this paper). Since the peak
optical flux from 
RS is proportional to $\sim t_d^{-1}$ (see equations \ref{fnuopta} \&
\ref{fnuoptb}), the flux will be reduced by $\sim$3-5 magnitudes and
therefore consistent with the observational upper limits of Kehoe et al.
(2001). In this case the early lightcurve should be rising, and subsequently
turn over to a steep decay at the deceleration time.
The most likely explanation for a typically faint optical flux
in our view is related to the low cooling frequency in the RS.
We see from the table 2 that cases with a bright optical flash have cooling 
frequency below the optical band at deceleration, in which case the lightcurve
should decline as $\sim t^{-3}$ and fade below the detection limit of 
14-15 magnitude in a few hundred seconds.

The observational situation (whether bright reverse shock emission is typical 
or not) should become clearer when \emph{Swift} is 
launched in September 2004. 

\section{Discussion}

We have explored the possibility that prompt $\gamma$-ray emission for a
selected sample of 10 long duration GRBs might arise in the external shocks. 
These bursts had good multiwavelength afterglow data and temporal coverage
which enabled Panaitescu and Kumar (2002) to determine their energy, 
jet opening angle, density of the surrounding medium, and microphysics 
parameters for the shock. 
We compared the observed peak flux and the peak frequency of the time
averaged spectrum during the $\gamma$-ray burst with the theoretical 
extrapolation of the afterglow emission
to the middle of the burst duration. 

The motivation for considering external shocks for the generation of 
$\gamma$-ray emission for these 10 bursts is that most of these bursts
are not highly variable, which is the primary reason for invoking
internal shocks. Moreover, the efficiency for the production of 
$\gamma$-rays for these bursts is found to be very high -- in excess
of 50\% -- which is difficult to understand in the internal shock models.
In the particular case of GRB 990123 the $\gamma$-ray pulse
width was of the same order as the time when the prompt optical emission from 
the reverse shock peaked i.e. the deceleration time. This means that the 
radius where $\gamma$-rays were generated was roughly the same as the 
deceleration radius, thereby suggesting a forward shock origin for gamma-rays.
Moreover, the low energy spectral slope $\alpha$ ($f_\nu\propto\nu^\alpha$)
for 990123 was 0.4, an observation that is difficult to understand in the 
internal shock models which have generally very low cooling frequency 
and therefore have $\alpha=-0.5$.

We find that it is not possible that the forward shock 
synchrotron afterglow emission, extrapolated back to prompt GRB duration,  
can explain the flux and peak frequency
of nine of these ten bursts; only in the case of 970508, which was a single
pulse FRED burst, does the extrapolation of the afterglow match up with the
$\gamma$-ray emission property. 
Moreover, it turns out that even when we take $\eBf$ (the 
energy fraction in magnetic field) and ISM density during the gamma-ray 
burst to be completely arbitrary, instead of the same as determined 
from afterglow modeling, we still cannot reconcile the gamma-ray observations
with the theoretical calculations for seven of the ten bursts in our
sample in the forward shock synchrotron emission model.
The reason for this is that the forward shock is highly radiative
at early times i.e., the cooling frequency during the gamma-ray burst 
is smaller than the synchrotron injection frequency, and therefore
the flux at the peak of the $\nu f_\nu$ spectrum is 
independent of the density of the surrounding medium and $\eBf$,
which happen to be the parameters with large uncertainty in 
afterglow modeling. 

Two of the bursts in the sample, 980519 \& 000418,  can be understood as
synchrotron emission in the forward shock provided that 
$\eBf\sim 1$ during the burst, a value that is larger by a 
factor$\sim 10^2$ than what we find during afterglow at $\sim 1$day;
it is unclear if this is a physically sensible solution.

We have also considered inverse-Compton scattering of synchrotron emission from
reverse or forward shock off of material in the forward and reverse shock
regions, and find that it is not possible to explain the $\gamma$-ray 
emission, except possibly for 980703, with a reasonable set of parameters. 
In particular, the
only solutions we found are when the circum-burst medium is taken to
have a wind like density profile with the density parameter about 
hundred times larger than the density of a typical Wolf-Rayet star wind 
and a few order of magnitude larger than the density determined from 
afterglow modeling. 

Adding large density clumps to the external medium only increases the
peak flux and frequency of $\nu f_{\nu}$ by a significant amount when
the density of the clump and ejecta are similar when they collide to
produce the $\gamma$-ray emission, in which case the early afterglow
emission is extremely bright and hard to miss.
Also, adding $e^{\pm}$ pairs to the ejecta
decreases the inverse Compton flux, unless the electron distribution
power law index is $\lta 2$, for any synchrotron-IC scattering scenario
considered. Thus, neither of these two possibilities seem likely to
explain the $\gamma$-ray emission properties of the bursts in our sample. 

So, we are therefore forced to conclude that there must be another way
to explain the GRB emission. The widely accepted internal shocks model might
be the solution. However, considering the problem of efficiency, special
cases of FREDs, and the problems for the internal shocks for 990123 
discussed in \S1, we feel it is prudent to explore other possible 
mechanisms such as the conversion of Poynting flux to radiation.

As for the reverse shock emission, we find that at least 50\% of these ten 
bursts have bright prompt optical flashes, 9-11th mag, provided that 
the shock parameters -- the energy fraction in electrons and magnetic 
field -- in the reverse shock are the same or larger than the value in the 
forward shock. However, five of these bursts have cooling frequencies below the optical band
and therefore the RS flux will decline very steeply with time (roughly
as $t^{-3}$) and could easily go undetected after a few deceleration
times. Those bursts with dimmer early emission generally have high cooling
frequencies, and are assumed to exhibit the expected $1/t^2$ fall
off.  Although they are dim, they may still be detected.  

If this sample of 10 bursts is representative of long duration GRBs 
then we expect very bright, rapidly fading ($\sim t^{-3}$),  prompt 
optical flashes accompanying many $\gamma$-ray bursts. The RS emission 
is particularly bright just above the synchrotron 
self-absorption frequency of $\sim 10^{-1}$ eV where we expect the 
observed flux to be of order a Jansky, and declining rapidly with time
since the cooling frequency is typically of the same order as the absorption
frequency. It is also possible that the deceleration time is much longer than
the GRB duration, which would reduce the predicted optical flux from the
reverse shock.  In this case we would expect to observe a dim, rising early
optical afterglow lightcurve, turning over to a steep descent
($t^{-2}$ or $t^{-3}$) after the deceleration time. 

These issues will be resolved in the \emph{Swift} era when we will have 
excellent early time coverage in the optical band for a few hours 
for many bursts. Future measurements of the early afterglow should enable 
us to determine if bright optical emission from the reverse shock is common, 
and thus determine the nature 
of the explosion, i.e. whether the explosion is baryonic, leptonic, or
Poynting-flux dominated.

\end{document}